# Adaptive output feedback tracking control for bilinear systems

Amir Reza Zare, Mahdi Aliyari shoorehdeli, Mehdi Tavan, Kamran Sabahi

*Corresponding author: Mehdi Tavan*

Amir Reza Zare is with the Department of Electrical Engineering, Science and research Branch, Islamic Azad University, Tehran, Iran.
Mahdi Aliyari shoorehdeli is with the Department of Electrical Engineering K. N. Toosi University of Technology, Tehran, Iran.
M. Tavan is with the Department of Electrical Engineering, Nour Branch, Islamic Azad University, Nour, Iran (e-mail: m.tavan@srbiau.ac.ir).
K. Sabahi is with the Department of Engineering Sciences, Faculty of Advanced Technologies, University of Mohaghegh Ardabili, Namin, Iran.

*Abstract* This paper deals with the trajectory tracking control problem for a class of bilinear systems with unmeasurable states and unknown parameters. Firstly, a full-information controller is suggested that guarantees global tracking under a persistency of excitation (PE) condition on the desired trajectories. Next, a model-based observer is designed for the system, which is further developed into an adaptive observer through dynamic regressor and mixing (DREM) parameter estimator. This enables global estimation under a weaker convergence condition where the regressor is PE. The estimated states and parameters are then replaced in the full-information controller, instead of their respective unavailable states and parameters, to construct the output feedback controller and its adaptive version. The proposed algorithm is applied to control lossless power factor precompensator (PFP) circuit with an unmeasurable input current and an unknown load conductance.

## I. Introduction

Bilinear systems are a type of nonlinear system that describes a wide range of physical and biological phenomena [1]. They can serve as a natural simplification of more complex nonlinear systems. There is a significant amount of literature dedicated to studying the intrinsic properties or stabilizing equilibrium points for these systems, as seen in [2], [3], [4], [5], [6], [7], and [8]. However, there are few results for designing controllers that ensure global tracking of admissible and differentiable trajectories, such as the PI controller proposed in [9]. To establish these results, an output signal must be constructed with respect to which the incremental model is passive. The tracking error will converge to zero if the output coefficient matrix, which relates the output error to the tracking error, is full rank for all time along the admissible trajectories.

This paper proposes a full-information controller that guarantees the global exponential stability of the tracking error under a PE assumption on the coefficient matrix, which is a strictly weaker condition than the full rank condition. Also, the stability of a Kalman-like observer is guaranteed for the system under a *light* observability assumption on the admissible control input. To cover parametric uncertainties of the system, the adaptive observer proposed in [10] has been developed based on the recent results in DREM-based estimator designs in [11] and [12]. This brings several advantages, including decoupling the parameter errors dynamic and ensuring global convergence under a non-square integrability condition of the extended regressor rather than traditional PE condition. Although this achievement applies an overparameterisation, the degree of freedom of the estimator is fewer than the generalized parameter estimation-based observer (GPEBO) such as the one presented in Proposition 1 of [22]. Also, under a square integrability condition on the output injection term in the observer, the proposed estimator ensures the convergence of the parameter estimation error despite the state estimation error convergence. The output feedback version of the proposed controller is constructed based on the foregoing observers.

An application of the proposed control algorithm is validated on a lossless model a single-phase full-bridge boost PFP. It is well known that this system is unobservable for zero control input while the admissible control input periodically passes through this point. It is shown that the proposed control algorithm gives a *certain* response to the system output feedback control problem.

The reminder of this paper is organized as follows. The problem formulation is presented in Section II. To enhance readability our response to the problem is split into three steps, given in three sections. A full-information controller is designed in Section III. For the unavailable states and parameters of the system, model based and adaptive observers are designed in Section IV. The stability of the certainty equivalent versions of the controller is analyzed in Section V. The theoretical results are applied to a physical example in section VI which includes simulation results.

## II. Problem formulation

Consider a class of nonlinear systems whose dynamics can be rewritten in suitable co-ordinates by equations of the following bilinear form

$$\dot{x} = [\mathcal{A}_0 + \textstyle\sum_{\mathfrak{i}=1}^{\mathfrak{u}} \mathcal{J}_{\mathfrak{i}} u_{\mathfrak{i}} - \mathcal{D}]x + \mathcal{B}_0(s)u + \mathcal{E}s, \tag{1}$$

$$y = \mathcal{C}^{\intercal}(u)x, \tag{2}$$

where $x \in \mathbb{R}^{\mathfrak{x}}$ is the state vector, $u \in \mathbb{R}^{\mathfrak{u}}$ is the control signal where $\mathfrak{u} \leq \mathfrak{x}$, $y \in \mathbb{R}^{\mathfrak{y}}$ is the output vector, $s \in \mathbb{R}^{\mathfrak{s}}$ is a vector of time-varying bounded signals, $\mathcal{A}_0 \in \mathbb{R}^{\mathfrak{x}\times\mathfrak{x}}$, $\mathcal{D} \in \mathbb{R}^{\mathfrak{x}\times\mathfrak{x}}$, $\mathcal{J}_{\mathfrak{i}} \in \mathbb{R}^{\mathfrak{x}\times\mathfrak{x}}$, and $\mathcal{E} \in \mathbb{R}^{\mathfrak{x}\times\mathfrak{s}}$ are real constant matrixes, $\mathcal{B}_0: \mathbb{R}^{\mathfrak{s}} \to \mathbb{R}^{\mathfrak{x}\times\mathfrak{u}}$ and $\mathcal{C}: \mathbb{R}^{\mathfrak{u}} \to \mathbb{R}^{\mathfrak{x}\times\mathfrak{y}}$ are globally Lipschitz.

*Definition 1* For some predefined desired output $y_d: \mathbb{R}_+ \to \mathbb{R}^{\mathfrak{y}}$, the bounded and differentiable signal $x_d: \mathbb{R}_+ \to \mathbb{R}^{\mathfrak{x}}$ is an *admissible* trajectory of the system (1)-(2), if it satisfies

$$\dot{x}_d = [\mathcal{A}_0 + \textstyle\sum_{\mathfrak{i}=1}^{\mathfrak{u}} \mathcal{J}_{\mathfrak{i}} u_{d\mathfrak{i}} - \mathcal{D}]x_d + \mathcal{B}_0(s)u_d + \mathcal{E}s, \tag{3}$$

$$y_d = \mathcal{C}^{\intercal}(u_d)x_d, \tag{4}$$

for some bounded control signal $u_d: \mathbb{R}_+ \to \mathbb{R}^{\mathfrak{u}}$ [9].

*Definition 2* For any arbitrary vectors $u \in \mathbb{R}^{\mathfrak{u}}$ let us define the mapping $\mathcal{J}: \mathbb{R}^{\mathfrak{u}} \to \mathbb{R}^{\mathfrak{x}\times\mathfrak{x}}$ as

$$\mathcal{J}(u) \coloneqq \textstyle\sum_{\mathfrak{i}=1}^{\mathfrak{u}} \mathcal{J}_{\mathfrak{i}} u_{\mathfrak{i}}, \tag{5}$$

and for any arbitrary vectors $x \in \mathbb{R}^{\mathfrak{x}}$ the mappings $\mathcal{I}: \mathbb{R}^{\mathfrak{x}} \to \mathbb{R}^{\mathfrak{x}\times\mathfrak{u}}$ as

$$\mathcal{I}(x) \coloneqq [\mathcal{J}_1 x, \dots, \mathcal{J}_{\mathfrak{u}} x], \tag{6}$$

Also, the coefficient matrixes

$$\mathcal{A}(u) \coloneqq \mathcal{A}_0 + \mathcal{J}(u), \tag{7}$$

$$\mathcal{B}(x, s) \coloneqq \mathcal{B}_0(s) + \mathcal{I}(x). \tag{8}$$

*Assumption 1* For the pair $(\mathcal{A}(u), \mathcal{D})$, there exist constant $\mathfrak{x} \times \mathfrak{x}$ matrixes $P = P^{\intercal} > 0$ and $\mathfrak{D} = \mathfrak{D}^{\intercal} \geq 0$ such that

$$P\mathcal{D} + \mathcal{D}^{\intercal}P = 2\mathfrak{D}\mathfrak{D}^{\intercal}, \tag{9}$$

$$P\mathcal{A}(u) = -\mathcal{A}^{\intercal}(u)P, \tag{10}$$

for any $u \in \mathbb{R}^{\mathfrak{u}}$ [9].

*Fact 1* From (10), it can be concluded that

$$x^{\intercal}P\mathcal{I}(x) = 0 \times u^{\intercal}. \tag{11}$$

*Assumption 2* There exist globally Lipschitz mappings $\Gamma: \mathbb{R}^{\mathfrak{u}} \to \mathbb{R}^{\mathfrak{x}\times\mathfrak{y}}$, $\mathcal{A}_{\sigma}, \mathcal{D}_{\sigma}: \mathbb{R}^{\mathfrak{u}} \to \mathbb{R}^{\mathfrak{x}\times\mathfrak{x}}$, and constant positive definite matrixes $P_{\sigma} = P_{\sigma}^{\intercal} \in \mathbb{R}^{\mathfrak{x}\times\mathfrak{x}}$ and $\mathfrak{D}_{\sigma} = \mathfrak{D}_{\sigma}^{\intercal} \in \mathbb{R}^{\mathfrak{y}\times\mathfrak{y}}$, such that

$$\mathcal{A}(u) - \mathfrak{d}\mathcal{D} - \Gamma(u)\mathcal{C}^{\intercal}(u) = \mathcal{A}_{\sigma}(u) - \mathcal{D}_{\sigma}(u), \tag{12}$$

with $\mathfrak{d} \in \{0, 1\}$, and the pair $\big(\mathcal{A}_{\sigma}(u), \mathcal{D}_{\sigma}(u)\big)$ satisfies (9)-(10) for $P = P_{\sigma}$ and $\mathfrak{D} = \mathcal{C}(u)\mathfrak{D}_{\sigma}$.

*Assumption 3* The origin of the dynamical system

$$\dot{\bar{x}} = [\mathcal{A}_{\sigma}(u) - \mathcal{D}_{\sigma}(u)]\bar{x}, \tag{13}$$

is globally exponentially stable (GES), uniformly in $u$, for all initial conditions $\bar{x}(0) \in \mathbb{R}^{\mathfrak{x}}$.

*Assumption 4* There exist known mappings $b: \mathbb{R}^{\mathfrak{y}} \times \mathbb{R}^{\mathfrak{u}} \times \mathbb{R}^{\mathfrak{s}} \to \mathbb{R}^{\mathfrak{x}}$ and $\Omega: \mathbb{R}^{\mathfrak{y}} \times \mathbb{R}^{\mathfrak{u}} \times \mathbb{R}^{\mathfrak{s}} \to \mathbb{R}^{\mathfrak{x}\times\mathfrak{p}}$ such that

$$\mathcal{B}_0(s)u + \mathcal{E}s - \mathcal{D}x = b(y, u, s) + \Omega(y, u, s)\theta, \tag{14}$$

where $\theta \in \mathbb{R}^{\mathfrak{p}}$ is a vector of unknown constant parameters.

*Assumption 5* The origin of the dynamical system

$$\dot{\bar{x}} = [\mathcal{A}_{\sigma}(u_d) - \mathcal{D}_{\sigma}(u_d)]\bar{x}, \tag{15}$$

is GES, uniformly in $u_d$, and for all initial conditions $\bar{x}(0) \in \mathbb{R}^{\mathfrak{x}}$.

*Fact 2* Since $u_d \in \mathcal{L}_{\infty}$, invoking the converse Lyapunov Theorem 4. 14 in [13], Assumption 5 guarantees the existence of a Lyapunov function $W_o: \mathbb{R}_{\geq 0} \times \mathbb{R}^{\mathfrak{x}} \to \mathbb{R}_{\geq 0}$ for the system (15) that satisfies the following inequalities

$$\mathfrak{c}_0|\bar{x}|^2 \leq W_o(t, \bar{x}) \leq \mathfrak{c}_1|\bar{x}|^2, \tag{16}$$

$$\nabla_t W_o + \nabla_{\bar{x}} W_o[\mathcal{A}_{\sigma}(u_d) - \mathcal{D}_{\sigma}] \leq -\mathfrak{c}_2|\bar{x}|^2, \tag{17}$$

$$|\nabla_{\bar{x}} W_o| \leq \mathfrak{c}_3|\bar{x}|, \tag{18}$$

for some positive constants $\mathfrak{c}_0$, $\mathfrak{c}_1$, $\mathfrak{c}_2$, and $\mathfrak{c}_3$.

Consider the system (1)-(2) under Assumptions 1-5, the control objective is to find mappings $\mathrm{H}: \mathbb{R}^{\mathfrak{e}} \times \mathbb{R}^{\mathfrak{y}} \times \mathbb{R}_+ \to \mathbb{R}^{\mathfrak{e}}$ and $\Xi: \mathbb{R}^{\mathfrak{e}} \times \mathbb{R}^{\mathfrak{y}} \times \mathbb{R}_+ \to \mathbb{R}^{\mathfrak{u}}$ such that the dynamic output feedback controller

$$\dot{\xi} = \mathrm{H}(\xi, y, t), \tag{19}$$

$$u = \Xi(\xi, y, t), \tag{20}$$

guarantees the trajectories of the closed-loop system remain bounded and

$$\lim_{t\to\infty} x = x_d, \tag{21}$$

for all initial conditions $x(0) \in \mathbb{R}^{x}$, $\xi(0) \in \mathbb{R}^{e}$, and all admissible trajectories $x_d$.

## III. Full-information controller

For the system, we define the error variables

$$\tilde{x} := x_d - x, \tag{22}$$

$$\tilde{u} := u_d - u. \tag{23}$$

Differentiating (22) and substituting (1), (3), (7), we get

$$\begin{aligned}\dot{\tilde{x}} &= \mathcal{A}(u_d)x_d - \mathcal{A}(u)x - \mathcal{D}\tilde{x} + \mathcal{B}_0(s)\tilde{u} \\ &= [\mathcal{A}(u) - \mathcal{D}]\tilde{x} + \mathcal{J}(\tilde{u})x_d + \mathcal{B}_0(s)\tilde{u} \\ &= [\mathcal{A}(u) - \mathcal{D}]\tilde{x} + \mathcal{B}(x_d, s)\tilde{u},\end{aligned} \tag{24}$$

where we have used

$$\mathcal{J}(\tilde{u})x_d = \mathcal{I}(x_d)\tilde{u}, \tag{25}$$

and (8) to get the last identity. Also, notice that, adding and subtracting the term $\mathcal{A}(u_d)x_d$ to the right-hand side of the first identity of (24), it can be shown, in a similar way, that

$$\dot{\tilde{x}} = [\mathcal{A}(u_d) - \mathcal{D}]\tilde{x} + \mathcal{B}(x, s)\tilde{u}. \tag{26}$$

*Proposition 1* Consider the system (1) verifying Assumption 1. Let

$$\begin{aligned}u &= u_{FI} \\ &:= u_d + K[\mathcal{B}_0^{\top}(s)Px_d - \mathcal{B}^{\top}(x_d, s)Px].\end{aligned} \tag{27}$$

Then for any positive definite matrix $K = K^{\top} \in \mathbb{R}^{u\times u}$, the trajectories of the closed-loop system is bounded and $\mathfrak{D}^{\top}\tilde{x}$, $\mathcal{B}^{\top}(x_d, s)P\tilde{x}$, and $\mathfrak{Q}^{\top}(x_d, s)\tilde{x}$ converge to zero where the real matrix $\mathfrak{Q}: \mathbb{R}^{x} \times \mathbb{R}^{s} \to \mathbb{R}^{x\times x}$ is such that

$$\mathfrak{Q}(x_d, s)\mathfrak{Q}^{\top}(x_d, s) = \mathfrak{D}\mathfrak{D}^{\top} + P\mathcal{B}(x_d, s)K\mathcal{B}^{\top}(x_d, s)P. \tag{28}$$

Assume that $\mathfrak{Q}(x_d, s) \in PE$, *i.e.*, there are $T, \beta, \alpha > 0$ such that the following *persistency of excitation* (PE) condition holds

$$\beta I \geq \int_{t_0}^{t_0+T} \mathfrak{Q}\big(x_d(\tau), s(\tau)\big)\mathfrak{Q}^{\top}\big(x_d(\tau), s(\tau)\big)d\tau \geq \alpha I, \tag{29}$$

for all $t_0 \in \mathbb{R}_+$. Then the system (24) has a uniformly globally exponentially stable (UGES) equilibrium at the origin provided that $\mathfrak{D}^{\top}\tilde{x} \equiv \mathcal{B}^{\top}(x_d, s)P\tilde{x} \equiv 0$ implies $\dot{\tilde{x}} \to 0$ as $t \to \infty$. Otherwise, there exists $\kappa > 0$ such that, for every $K \geq \kappa_\circ I$, the origin of (24) is globally exponentially stable.

*Proof 1* To begin with, note that

$$\begin{aligned}u_{FI} &= u_d + K[\mathcal{B}_0^{\top}(s)Px_d - \mathcal{B}^{\top}(x_d, s)Px] \\ &= u_d + K\mathcal{B}^{\top}(x_d, s)P\tilde{x},\end{aligned} \tag{30}$$

where (11) of Fact 1 has been used to get the second identity. Replacing (30) in (23), and its result in (24), yields the following *linear time-varying* system

$$\dot{\tilde{x}} = [\mathcal{A}(u_{FI}) - \mathcal{D} - \mathcal{B}(x_d, s)K\mathcal{B}^{\top}(x_d, s)P]\tilde{x}, \tag{31}$$

which confirms the origin is an invariance set for the tracking error dynamics. Now, consider the Lyapunov function

$$V_c(\tilde{x}) = \frac{1}{2}|\mathfrak{P}^{\top}\tilde{x}|^2, \tag{32}$$

where the real matrix $\mathfrak{P} \in \mathbb{R}^{x\times x}$ is the square root of $P$ such that $\mathfrak{P}\mathfrak{P}^{\top} = P$. Differentiating (32) along the trajectories (31), and using (9) and (10), yields

$$\begin{aligned}\dot{V}_c &= -|\mathfrak{Q}^{\top}(x_d, s)\tilde{x}|^2 \\ &= -|\mathfrak{D}^{\top}\tilde{x}|^2 - |\mathfrak{K}^{\top}\mathcal{B}^{\top}(x_d, s)P\tilde{x}|^2,\end{aligned} \tag{33}$$

where the real matrix $\mathfrak{K} \in \mathbb{R}^{u\times u}$ is the unique square root of $K$ in other words $\mathfrak{K}\mathfrak{K}^{\top} = \mathfrak{K}^2 = K$. This implies that $V_c \in \mathcal{L}_\infty$, and consequently $\tilde{x} \in \mathcal{L}_\infty$ and $\mathfrak{Q}^{\top}(x_d, s)\tilde{x} \in \mathcal{L}_2$, $\mathfrak{D}^{\top}\tilde{x} \in \mathcal{L}_2$, $\mathcal{B}^{\top}(x_d, s)P\tilde{x} \in \mathcal{L}_2$. Now, notice that due to $x_d$ and $u_d$ belonging to $\mathcal{L}_\infty$ we get the control signal $u$, given by (27), belongs to $\mathcal{L}_\infty$. This implies $\dot{\tilde{x}} \in \mathcal{L}_\infty$ concerning (24). Since $x_d$ and $\mathcal{B}_0(s)$ are continuous and bounded, we get $\dot{\mathfrak{Q}} \in \mathcal{L}_\infty$. This implies that the square-integrable signal $\mathfrak{Q}^{\top}(x_d, s)\tilde{x}$ has a bounded time-derivative. Hence, the uniform asymptotic convergence of $\mathfrak{Q}^{\top}(x_d, s)\tilde{x}$, $\mathfrak{D}^{\top}\tilde{x}$, and $\mathcal{B}^{\top}(x_d, s)P\tilde{x}$ to zero can be concluded from the generalization of Barbalat's Lemma [14]. This in turn implies from Lemma 1 in [15] that $\tilde{x}$ asymptotically converges to zero provided that $\mathfrak{Q} \in PE$ and $\dot{\tilde{x}} \to 0$. The proof of the claim is completed by noting that, for linear time-varying systems, uniform asymptotic stability confirms exponential stability by Theorem 4.11 in [13]. Now, let

$$K = \mathscr{k} K_0, \tag{34}$$

where $\mathscr{k} > 0$, then (31) can be rewritten as

$$\begin{aligned}\dot{\tilde{x}} &= -\mathscr{k} P^{-1}(\mathfrak{D}\mathfrak{D}^\top + P\mathcal{B}K_0\mathcal{B}^\top P)\tilde{x} + (\mathcal{A}(u_{FI}) - \mathcal{D} + \mathscr{k} P^{-1}\mathfrak{D}\mathfrak{D}^\top)\tilde{x} \\ &= -\mathscr{k}\mathfrak{P}^{-\top}\mathfrak{Q}_0(x_d, s)\mathfrak{Q}_0^\top(x_d, s)\mathfrak{P}^\top\tilde{x} + \mathfrak{P}^{-\top}d(\tilde{x}, x_d, u_d, s)\mathfrak{P}^\top\tilde{x},\end{aligned} \tag{35}$$

with

$$\mathfrak{Q}_0(x_d, s)\mathfrak{Q}_0^\top(x_d, s) = \mathfrak{P}^{-1}(\mathfrak{D}\mathfrak{D}^\top + P\mathcal{B}(x_d, s)K_0\mathcal{B}^\top(x_d, s)P)\mathfrak{P}^{-\top}, \tag{36}$$

and

$$\begin{aligned}d(\tilde{x}, x_d, u_d, s) &= \mathfrak{P}^\top(\mathcal{A}(u_{FI}) - \mathcal{D} + \mathscr{k} P^{-1}\mathfrak{D}\mathfrak{D}^\top)\mathfrak{P}^{-\top} \\ &= \mathfrak{P}^\top(\mathcal{A}(u_d) - \mathcal{D} + \mathcal{J}(\tilde{x})K\mathcal{B}^\top(x_d, s)P + \mathscr{k} P^{-1}\mathfrak{D}\mathfrak{D}^\top)\mathfrak{P}^{-\top},\end{aligned} \tag{37}$$

where (25) and (30) have been used to get the above identity. The change of coordinates

$$w := \mathfrak{P}^\top\tilde{x}, \tag{38}$$

transforms the above system into

$$\dot{w} = -\mathscr{k}\mathfrak{Q}_0(x_d, s)\mathfrak{Q}_0^\top(x_d, s)w + d(\tilde{x}, x_d, u_d, s)w. \tag{39}$$

It is clear that $\mathfrak{Q} \in PE$ confirms that $\mathfrak{Q}_0 \in PE$ due to $\mathfrak{P}$ is full rank. This implies that the unforced system

$$\dot{w} = -\mathscr{k}\mathfrak{Q}_0(x_d, s)\mathfrak{Q}_0^\top(x_d, s)w, \tag{40}$$

has a UGES equilibrium at zero [16]. Consequently, similar arguments to Fact 2 can be represented for the above system. Define

$$\mathcal{P}(t) := \int_t^{+\infty} \Psi(\tau, t)\, \Psi^\top(\tau, t) d\tau, \tag{41}$$

where $\Psi(\tau, t)$ is the state transition matrix of the system (40). It is worth pointing out that $\mathcal{P}(t)$ is positive definite, bounded, and satisfies the lower bound [16]

$$\|\mathcal{P}(t)\| \geq \frac{1}{2}\mathscr{k}^{-1}\rho^{-1}, \tag{42}$$

where $\rho := \sup\|\mathfrak{Q}_0(x_d, s)\|^2$ that is positive and bounded due to $x_d,\ s \in \mathcal{L}_\infty$. From Theorem 4.12 in [13], the time derivative of the Lyapunov function

$$W_c(w) = w^\top\mathcal{P}(t)w, \tag{43}$$

along the trajectories (40) is

$$\dot{W}_c = -|w|^2. \tag{44}$$

Consider the candidate Lyapunov function

$$V(w) = W_c + \sigma V_c \tag{45}$$

with $\sigma > 0$, whose time derivative along the trajectories (39) is given by

$$\begin{aligned}\dot{V} &= -|w|^2 + 2w^\top\mathcal{P}dw + \sigma\dot{V}_c \\ &= -|w|^2 - \sigma|\mathfrak{Q}^\top\mathfrak{P}^{-\top}w|^2 + 2w^\top\mathcal{P}dw,\end{aligned} \tag{46}$$

where (33) has been rephrased in terms of $w$ to get the second identity. An upper bound for the last term can be obtained by

$$\begin{aligned}&w^\top\mathcal{P}dw \leq \\ &\|\mathcal{P}\|\|\mathfrak{P}^\top(\mathcal{A}(u_d) - \mathcal{D})\mathfrak{P}^{-\top}\||w|^2 + \mathscr{k}|\mathfrak{D}^\top\mathfrak{P}^{-\top}w|\|\mathfrak{P}^{-1}\mathfrak{D}\|\|\mathcal{P}\||w| + \\ &|\mathfrak{K}^\top\mathcal{B}^\top(x_d, s)P\mathfrak{P}^{-\top}w|\|\mathcal{J}(\tilde{x})\mathfrak{K}\|\|\mathcal{P}\||w|.\end{aligned} \tag{47}$$

Using Young inequality, the last two terms above satisfy

$$2\mathscr{k}|\mathfrak{D}^\top\mathfrak{P}^{-\top}w|\|\mathfrak{P}^{-1}\mathfrak{D}\|\|\mathcal{P}\||w| \leq \|\mathfrak{P}^{-1}\mathfrak{D}\|^2\|\mathcal{P}\||\mathfrak{D}^\top\mathfrak{P}^{-\top}w|^2 + \|\mathcal{P}\||w|^2, \tag{48}$$

$$2|\mathfrak{K}^\top\mathcal{B}^\top(x_d, s)P\mathfrak{P}^{-\top}w|\|\mathcal{J}(\tilde{x})\mathfrak{K}\|\|\mathcal{P}\||w| \leq \|\mathcal{J}(\tilde{x})\mathfrak{K}\|^2\|\mathcal{P}\||\mathfrak{K}^\top\mathcal{B}^\top(x_d, s)P\mathfrak{P}^{-\top}w|^2 + \|\mathcal{P}\||w|^2. \tag{49}$$

Notice that, from (32)-(33), $\dot{V}_c \leq 0$ implies

$$|\tilde{x}(t)| \leq \tilde{x}_M := \sqrt{p_M/p_m}|\tilde{x}(0)|, \tag{50}$$

where $p_m$ and $p_M$ are the minimum and maximum eigenvalues of $P$. Since $\mathcal{P}(t)$ is bounded, we can choose $\sigma = 2\sup(\|\mathcal{P}\|) \times \max\{\|\mathfrak{P}^{-1}\mathfrak{D}\|^2, \|\mathcal{J}(\tilde{x}_M)\mathfrak{K}\|^2\}$. Now, recalling (33) and using (47)-(49), $\dot{V}$ can be bounded by

$$\dot{V} \leq -\big(1 - 2\|\mathcal{P}\|(1 + \|\mathfrak{P}^\top(\mathcal{A}(u_d) - \mathcal{D})\mathfrak{P}^{-\top}\|)\big)|w|^2. \tag{51}$$

Now, invoking (42) and choosing

$$\mathscr{k} > \kappa_\diamond := \rho^{-1}\sup(1 + \|\mathfrak{P}^\top(\mathcal{A}(u_d) - \mathcal{D})\mathfrak{P}^{-\top}\|), \tag{52}$$

the right-hand side of (51) is negative definite and the proof is completed.

*Remark 1* The second identity in (30) reminds us of the control law proposed in [17] for the regulation problem of DC-DC buck and boost converters. The PI controller proposed in [9] and the controller proposed in (27) can be considered as an extension of [17] for the trajectory tracking problem. For the system with $\mathcal{B}_0(s) = 0$, (27) is equivalent to the PI controller with zero integral gain. The global asymptotic stability of the tracking error dynamics is ensured in [17] for $\mathfrak{Q}(x_d, s)$ of full rank for all time, while in Proposition 1 the global exponential stability of the dynamics is guaranteed under a strictly weaker condition, i.e., $\mathfrak{Q}(x_d, s) \in PE$.

## IV. OBSERVER DESIGN

The following proposition proposes a Kalman-like observer which requires the parameters of the system. To begin with, define the state estimation error as

$$\tilde{x} := x - \hat{x}, \tag{53}$$

where $\hat{x} \in \mathbb{R}^{\mathfrak{x}}$ is the estimated state.

*Proposition 2* Consider system (1)-(2) verifying Assumption 2, in conjunction with the observer

$$\dot{\hat{x}} = [\mathcal{A}(u) - \mathcal{D} - \Gamma(u)\mathcal{C}^\top(u)]\hat{x} + \Gamma(u)y + \mathcal{B}_0(s)u + \mathcal{E}s, \tag{54}$$

where $\Gamma(u)$ is specified in Assumption 2. Then $\tilde{x} = 0$ is globally stable (GS) and $\mathcal{C}^\top(u)\tilde{x} \in \mathcal{L}_2$, uniformly in $u$. Moreover, if Assumption 3 holds, then $\hat{x}$ is an GE converging estimate of $x$, uniformly in $u$.

*Proof 2* Dedifferentiating (53) and replacing (1) and (54) therein, yields the following error dynamics

$$\begin{aligned}\dot{\tilde{x}} &= [\mathcal{A}(u) - \mathcal{D} - \Gamma(u)\mathcal{C}^\top(u)]\tilde{x} \\ &= [\mathcal{A}_o(u) - \mathcal{D}_o(u)]\tilde{x},\end{aligned} \tag{55}$$

where (12) has been used to get the second equality. Now, notice that, invoking Assumption 2, the time derivative of the Lyapunov candidate function

$$V_o(\tilde{x}) = \frac{1}{2}\tilde{x}^\top P_o \tilde{x}, \tag{56}$$

along the trajectories (55) is

$$\dot{V}_o(\tilde{x}) = -|\mathfrak{D}_o^\top \mathcal{C}^\top(u)\tilde{x}|^2, \tag{57}$$

which implies the origin of (55) is GS, uniformly in $u$, and $\mathcal{C}^\top(u)\tilde{x} \in \mathcal{L}_2$ due to $\mathfrak{D}_o$ being full rank. The proof is completed by invoking Assumption 3.

To cover parametric uncertainties, an adaptive observer is designed for a class of bilinear systems, which verifies Assumption 4, in the next step. To facilitate transforming the system dynamics to a proper adaptive observer form, the following filtered transformation is considered

$$z := x - \Upsilon(t)\theta, \tag{58}$$

where $\Upsilon: \mathbb{R}_+ \to \mathbb{R}^{\mathfrak{x}\times\mathfrak{p}}$ is an auxiliary filter whose dynamics are yet to be specified. Using (14), the system (1)-(2) can be represented in $z$-coordinate as

$$\dot{z} = \mathcal{A}(u)z + b(y, u, s) + [\mathcal{A}(u)\phi + \Omega_o(y, u, s) - \dot{\Upsilon}]\,\theta, \tag{59}$$

$$y = \mathcal{C}^\top(u)z + \mathcal{C}^\top(u)\Upsilon(t)\theta. \tag{60}$$

Inspired by [10], [11] and [12], an adaptive observer is proposed in the following proposition. Before that, let us define the state and parameter estimation errors as

$$\tilde{z} := z - \hat{z}, \tag{61}$$

$$\tilde{\theta} := \theta - \hat{\theta}, \tag{62}$$

where $\hat{z} \in \mathbb{R}^{\mathfrak{x}}$ and $\hat{\theta} \in \mathbb{R}^{\mathfrak{p}}$ are the estimated states and parameters.

*Proposition 3* Consider the system (1)-(2) verifying Assumption 4, in conjunction with the observer

$$\dot{\hat{z}} = [\mathcal{A}(u) - \Gamma(u)\mathcal{C}^\top(u)]\hat{z} + \Gamma(u)y + b(y, u, s), \tag{63}$$

$$\dot{\hat{\theta}} = \Lambda\,\mathrm{adj}[\Phi^\top(t)\Phi(t)]\,\Phi^\top(t)\big(\mathcal{Y} - \Phi(t)\hat{\theta}\big), \tag{64}$$

with

$$\dot{\Upsilon} = [\mathcal{A}(u) - \Gamma(u)\mathcal{C}^\top(u)]\Upsilon + \Omega(y, s), \tag{65}$$

$$\dot{\mathcal{Y}} = -\mathrm{T}\mathcal{Y} + \Upsilon^\top(t)\mathcal{C}(u)(y - \mathcal{C}^\top(u)\hat{z}), \tag{66}$$

$$\dot{\Phi} = -\mathrm{T}\Phi + \Upsilon^\top(t)\mathcal{C}(u)\mathcal{C}^\top(u)\Upsilon(t), \tag{67}$$

where $\Gamma$ is specified in Assumption 2, the $\mathfrak{p}\times\mathfrak{p}$ constant matrixes $\Lambda > 0$ and $\mathrm{T} > 0$ are diagonal, $\Upsilon(0) \in \mathbb{R}^{\mathfrak{x}\times\mathfrak{p}}$, $\mathcal{Y}(0) \in \mathbb{R}^{\mathfrak{p}\times\mathfrak{y}}$, the $\mathfrak{p}\times\mathfrak{p}$ matrix $\Phi(0) > 0$, and adj[.] denotes the adjugate matrix. If Assumption 2 holds, then $\tilde{z} = 0$ is GS, uniformly in $u$. In addition, if Assumption 3 holds, then $\hat{z}$ is a GE converging estimate of $z$ and $\mathcal{C}^\top(u)\tilde{z} \in \mathcal{L}_2$, uniformly in $u$. Moreover, if $u \in \mathcal{L}_\infty$ and $\Omega \in \mathcal{L}_\infty$, then $\Upsilon$ is bounded. If in addition $\det[\Phi(t)] \notin \mathcal{L}_2$, then $\hat{\theta}$ is an uniformly globally asymptotically (UGA) converging estimate of $\theta$. Furthermore, if $\Upsilon^\top(t)\mathcal{C}(u(t)) \in$ $PE$, then $\hat{\theta}$ is a uniformly globally exponentially (UGE) converging estimate of $\theta$. If $\Upsilon^\top(t)\mathcal{C}(u(t)) \in PE$ and only Assumption 2 holds, then $\hat{\theta}$ is a UGA converging estimate of $\theta$.

*Proof 3* To begin with, note that, differentiating (61) and using (59), (63), and (65), yields

$$\begin{aligned}\dot{\bar{z}} &= [\mathcal{A}(u) - \Gamma(u)\mathcal{C}^\top(u)]\bar{z} \\ &= [\mathcal{A}_o(u) - \mathcal{D}_o(u)]\bar{z}.\end{aligned} \quad (68)$$

Invoking Assumption 2, the time derivative of the Lyapunov function

$$V_o(\bar{z}) = \frac{1}{2}\bar{z}^\top P_o \bar{z}, \quad (69)$$

along the trajectories (68) is

$$\dot{V}_o(\bar{z}) = -|\mathfrak{D}_o^\top \mathcal{C}^\top(u)\bar{z}|^2, \quad (70)$$

which implies $V_o \in \mathcal{L}_\infty$, $\mathcal{C}^\top(u)\bar{z} \in \mathcal{L}_2$, and the system (68) has a GS equilibrium at the origin, uniformly in $u$, which is also GES with respect to Assumption 3. Also, from Assumption 3 and $u \in \mathcal{L}_\infty$, we obtain (65) is input-to-state stable by Lemma 4.6 in [13]. Consequently, $\Omega \in \mathcal{L}_\infty$ implies $\Upsilon \in \mathcal{L}_\infty$. Now, multiplying (60) by $\Upsilon^\top\mathcal{C}$ on the left, yields

$$\Upsilon^\top(t)\mathcal{C}(u)(y - \mathcal{C}^\top(u)z) = \Upsilon^\top(t)\mathcal{C}(u)\mathcal{C}^\top(u)\Upsilon(t)\theta. \quad (71)$$

This implies, by linearity, the matrixes $\mathcal{Y}$ and $\Phi$, generated by (66) and (67), satisfy

$$\mathcal{Y}(t) = \Phi(t)\theta + \epsilon(t), \quad (72)$$

where $\epsilon(t)$ is generated by

$$\dot{\epsilon} = -\mathrm{T}\epsilon - \Upsilon^\top(t)\mathcal{C}(u)\mathcal{C}^\top(u)\bar{z}, \quad (73)$$

which decays to zero exponentially due to $\bar{z}$ has this property, and $\Upsilon$ and $\mathcal{C}$ are bounded. Multiplying (72) by $\Lambda\,\mathrm{adj}[\Phi^\top\Phi]\,\Phi^\top$ on the left, yields

$$\Lambda\,\mathrm{adj}[\Phi^\top(t)\Phi(t)]\,\Phi^\top(t)\big(\mathcal{Y}(t) - \epsilon(t)\big) = \det{}^2\big(\Phi(t)\big)\Lambda\theta, \quad (74)$$

where the facts that for any matrix $N \in \mathbb{R}^{\mathfrak{p}\times\mathfrak{p}}$, we have that $\det(N)\,I = \mathrm{adj}(N)\,N = N\,\mathrm{adj}(N)$, $\det(N^\top N) = \det(N^\top)\det(N)$, and $\det(N^\top) = \det(N)$, have been used respectively to get the equality. Differentiating (62), replacing (64) therein, and using (74), we get

$$\dot{\bar{\theta}}_i = -\lambda_i \det{}^2\big(\Phi(t)\big)\,\bar{\theta}_i + \varepsilon_i(t), \quad (75)$$

for $i = 1, \dots, \mathfrak{p}$, where $\varepsilon_i(t)$ is picked from

$$\varepsilon(t) = -\Lambda\,\mathrm{adj}[\Phi^\top(t)\Phi(t)]\,\Phi^\top(t)\epsilon(t), \quad (76)$$

which decays to zero exponentially and satisfies $\varepsilon_i \in \mathcal{L}_1$. Under $\det[\Phi(t)] \notin \mathcal{L}_2$ condition, we can apply Lemma 1 in [18] to the parameter estimation error dynamics (75) to conclude the asymptotic convergence claim. Now, notice that $\Upsilon^\top\mathcal{C} \in PE$ implies $\Phi \in PE$ [11]. As a result, the unperturbed part of (75) is UGES by Theorem 1 in [11]. For the square matrix $\Phi$, $\Phi \in PE$ implies $\det[\Phi(t)] \notin \mathcal{L}_2$, consequently, the trajectories of (75) remain bounded. Now, UGES of the system (75) can be concluded from Proposition 2.3 in [19] by noting that the perturbation term $\varepsilon_i(t)$ tends to zero exponentially.

From $\Upsilon^\top\mathcal{C} \in PE$ it can be concluded that $\Upsilon^\top\mathcal{C} \in \mathcal{L}_\infty$ and then $\Phi \in \mathcal{L}_\infty$. As we mentioned above, the origin of the *nominal* system

$$\dot{\bar{\theta}} = -\Lambda\det{}^2\big(\Phi(t)\big)\,\bar{\theta}. \quad (77)$$

is UGES under $\Upsilon^\top\mathcal{C} \in PE$. Consequently, similar arguments to (41)-(44) can be repeated for the system. To this end, let

$$\Lambda = \lambda\Lambda_0^2, \quad (78)$$

where $\lambda > 0$ and $\Lambda_0 > 0$ are constant scalar and diagonal matrices, respectively. Define

$$\Theta(t) := \int_t^{+\infty} \Psi(\tau, t)\Psi^\top(\tau, t)d\tau, \quad (79)$$

where this time $\Psi(\tau, t)$ is the state transition matrix of the system (77) and we know that $\Theta(t)$ is positive definite, bounded, and satisfies [16]

$$\|\Theta(t)\| \geq \frac{1}{2}\lambda^{-1}\vartheta^{-1}, \quad (80)$$

where $\vartheta := \sup\|\Lambda_0\det(\Phi(t))\|^2$ that is positive and bounded due to $\Phi \in \mathcal{L}_\infty$. Also, from Theorem 4.12 in [13], the time derivative of the Lyapunov function

$$W_e(\bar{\theta}) = \bar{\theta}^\top\Theta(t)\bar{\theta}, \quad (81)$$

along the trajectories (94) is

$$\dot{W}_e = -|\bar{\theta}|^2. \quad (82)$$

Now, notice that if Assumption 2 holds, the time derivative of the Lyapunov function

$$V_e(t, \epsilon, \bar{z}, \bar{\theta}) = V_o(\bar{z}) + \frac{1}{2}\sigma_e|\epsilon|^2 + W_e(t, \bar{\theta}), \quad (83)$$

with $\sigma_e > 0$, along the trajectories (68), (73), and (75)-(76), satisfies

$$\dot{V}_e = -|\mathfrak{D}_\sigma^\top \mathcal{C}^\top(u)\bar{z}|^2 - \tfrac{1}{2}\sigma_e|\mathfrak{T}^\top\epsilon|^2 - \left|\bar{\theta}\right|^2 + \epsilon^\top\Upsilon^\top(t)\mathcal{C}(u)\mathcal{C}^\top(u)\bar{z} + \bar{\theta}^\top\Theta(t)\varepsilon(t). \tag{84}$$

where $\mathfrak{T} \in \mathbb{R}^{p\times p}$ is the square root of T. Using the inequalities

$$2\epsilon^\top\Upsilon^\top(t)\mathcal{C}(u)\mathcal{C}^\top(u)\bar{z} \leq |\mathfrak{D}_\sigma^\top \mathcal{C}^\top(u)\bar{z}|^2 + \mathfrak{a}_0|\mathfrak{T}^\top\epsilon|^2, \tag{85}$$

$$2\,\bar{\theta}^\top\Theta(t)\varepsilon(t) \leq \left|\bar{\theta}\right|^2 + \mathfrak{a}_1|\mathfrak{T}^\top\epsilon|^2, \tag{86}$$

where $\mathfrak{a}_0 := \sup(\|\mathfrak{T}^{-1}\Upsilon^\top(t)\mathcal{C}(u)\mathfrak{D}_\sigma^{-\top}\|^2)$ and $\mathfrak{a}_1 := \sup(\|\Lambda\,\mathrm{adj}[\Phi^\top(t)\Phi(t)]\,\Phi^\top(t)\|^2)$. Now, by choosing $\sigma_e = 1 + \mathfrak{a}_0 + \mathfrak{a}_1$, we have

$$\dot{V}_e \leq -\tfrac{1}{2}|\mathfrak{D}_\sigma^\top \mathcal{C}^\top(u)\bar{z}|^2 - \tfrac{1}{2}|\mathfrak{T}^\top\epsilon|^2 - \tfrac{1}{2}\left|\bar{\theta}\right|^2, \tag{87}$$

which implies $V_e$, $\bar{z}$, $\epsilon$, $\bar{\theta}$ belong to $\mathcal{L}_\infty$ and, consequently, we have $\mathcal{C}^\top(u)\bar{z}$, $\epsilon$, $\bar{\theta}$ belong to $\mathcal{L}_2$. As a result, from (76) we get $\varepsilon \in \mathcal{L}_\infty$ due to $\epsilon \in \mathcal{L}_\infty$ and $\Phi \in \mathcal{L}_\infty$. This, from (75), yields $\dot{\bar{\theta}} \in \mathcal{L}_\infty$. Now, UGAS of $\bar{\theta} = 0$ can be concluded by Lemma 1 in [14]. This completed the proof.

*Remark 2* As shown in [20] and [21], the convergence condition $\det[\Phi(t)] \notin \mathcal{L}_2$ is a strictly weaker condition than $\Upsilon^\top(t)\mathcal{C}(u) \in PE$ which is used in [10]. Although this achievement applies an overparameterisation, the degree of freedom of the estimator is fewer than the generalized parameter estimation-based observer (GPEBO) such as the one presented in Proposition 1 of [22]. The last claim of Proposition 3 shows that under assumption $\Upsilon^\top\mathcal{C} \in PE$, $\bar{\theta}$ can tend to zero for state estimation errors belonging to the set $\mathcal{C}^\top\bar{z} \in \mathcal{L}_2 \cap \mathcal{L}_\infty$ which includes even non-zero $\bar{z}$.

## V. Output Feedback Controller Design

The full-information controller of Proposition 1 can be rewritten as a function of $x$ and $\theta$ as following

$$u_{FI} = u_{FI}(x, \theta). \tag{88}$$

An output feedback form of the control law can be obtained by replacing the unavailable states with their estimates, provided by the observer of Proposition 2.

*Proposition 4* Consider the system (1)-(2) verifying Assumptions 1-2 and 5, in conjunction with the observer (54). If the origin of the system (31) is UGES for all $K = K^\top > 0$, then there exists a sufficiently small constant $\kappa_\star$ such that the system in closed loop with $u = u_{FI}(\hat{x}, \theta)$ is UGES for all $\kappa_\star I > K > 0$.

*Proof 4* Replacing $u = u_{FI}(\hat{x}, \theta)$ in (23) and using (30), we get

$$\tilde{u} = -K\mathcal{B}^\top(x_d, s)P(\tilde{x} + \bar{x}). \tag{89}$$

Substituting the above equation in (24) and invoking the property (25), yields the following tracking error dynamics

$$\begin{aligned}\dot{\tilde{x}} &= [\mathcal{A}(u) - \mathcal{D} - \mathcal{B}(x_d, s)K\mathcal{B}^\top(x_d, s)P]\tilde{x} - \mathcal{B}(x_d, s)K\mathcal{B}^\top(x_d, s)P\bar{x},\\ &= [\mathcal{A}(u_{FI}) - \mathcal{D} - \mathcal{B}(x_d, s)K\mathcal{B}^\top(x_d, s)P]\tilde{x} + \big(\mathcal{I}(\tilde{x}) - \mathcal{B}(x_d, s)\big)K\mathcal{B}^\top(x_d, s)P\bar{x},\end{aligned} \tag{90}$$

which can be seen as the perturbed form of the *nominal* system (31) with perturbation $\bar{x}$. From Proposition 2, $\bar{x} \in \mathcal{L}_\infty$, uniformly in $u$. Now, notice that (55) can be rewritten in terms of $u_d$ and $\tilde{u}$, as following

$$\begin{aligned}\dot{\bar{x}} &= [\mathcal{A}_\sigma(u_d) - \mathcal{D}_\sigma]\bar{x} - \mathcal{J}_\sigma(\tilde{u})\bar{x}\\ &= [\mathcal{A}_\sigma(u_d) - \mathcal{D}_\sigma]\bar{x} - \mathcal{I}_\sigma(\bar{x})\tilde{u},\end{aligned} \tag{91}$$

with the matrixes $\mathcal{J}_\sigma$ and $\mathcal{I}_\sigma$ which play the role $\mathcal{J}$ and $\mathcal{I}$ for $\mathcal{A}_\sigma(u)$. The above dynamics can be seen as the perturbed form of the *nominal* system (15) with perturbation $\tilde{u}$. Now, a Lyapunov candidate function for the closed-loop system can be found as

$$V_{cl}(t, \bar{x}, \tilde{x}) = W_o(t, \bar{x}) + V_c(\tilde{x}), \tag{92}$$

where $W_o$ and $V_c$ are given by (16)-(18) and (32), respectively. The time derivative of (92) along the trajectories (90) and (91) satisfies

$$\begin{aligned}\dot{V}_{cl}(t) \leq &-\mathfrak{c}_2|\bar{x}|^2 - |\mathfrak{D}^\top\tilde{x}|^2 - |\mathfrak{K}^\top\mathcal{B}^\top(x_d, s)P\tilde{x}|^2\\ &+\mathfrak{c}_3|\bar{x}||\mathcal{I}_\sigma(\bar{x})\tilde{u}| + |\mathfrak{K}^\top\mathcal{B}^\top(x_d, s)P\tilde{x}||\mathfrak{K}^\top\mathcal{B}^\top(x_d, s)P\bar{x}|,\end{aligned} \tag{93}$$

where (33) and (17)-(18) have been used. Notice that, (56) and (57) imply

$$|\bar{x}(t)| \leq \bar{x}_M := \sqrt{p_{\sigma M}/p_{\sigma m}}|\bar{x}(0)|, \tag{94}$$

where $p_{\sigma m}$ and $p_{\sigma M}$ are the minimum and maximum eigenvalues of $P_\sigma$. Replacing (89) in (93) and using the following inequalities

$$\mathfrak{c}_3|\mathcal{I}_\sigma(\bar{x})K\mathcal{B}^\top(x_d, s)P\bar{x}| \leq k\mathfrak{b}_0|\bar{x}|, \tag{95}$$

$$\mathfrak{c}_3|\bar{x}||\mathcal{I}_\sigma(\bar{x})K\mathcal{B}^\top(x_d, s)P\tilde{x}| \leq k\mathfrak{b}_1|\bar{x}|^2 + \tfrac{1}{4}|\mathfrak{K}^\top\mathcal{B}^\top(x_d, s)P\tilde{x}|^2, \tag{96}$$

$$|\mathfrak{K}^\top\mathcal{B}^\top(x_d, s)P\tilde{x}||\mathfrak{K}^\top\mathcal{B}^\top(x_d, s)P\bar{x}| \leq k\mathfrak{b}_2|\bar{x}|^2 + \tfrac{1}{4}|\mathfrak{K}^\top\mathcal{B}^\top(x_d, s)P\tilde{x}|^2, \tag{97}$$

with $\mathfrak{b}_0 := \max(\mathfrak{c}_3\|\mathcal{J}_\sigma(\bar{x}_M)K_0\mathcal{B}^\top(x_d,s)P\|)$, $\mathfrak{b}_1 := \max(\mathfrak{c}_3\|K_0\|\|\mathcal{J}_\sigma(\bar{x}_M)\|^2)$, $\mathfrak{b}_2 := \max(\|K_0\|\|\mathcal{B}^\top(x_d,s)P\|^2)$, we get

$$\dot{V}_{cl}(t) \le -(\mathfrak{c}_2 - k(\mathfrak{b}_0+\mathfrak{b}_1+\mathfrak{b}_2))|\bar{x}|^2 - |\mathfrak{D}^\top\tilde{x}|^2 - \frac{1}{2}|\mathfrak{K}^\top\mathcal{B}^\top(x_d,s)P\tilde{x}|^2, \tag{98}$$

which is negative for

$$k < \kappa_\star := \mathfrak{c}_2/(\mathfrak{b}_0+\mathfrak{b}_1+\mathfrak{b}_2). \tag{99}$$

As a result, for $K < \kappa_\star I$, we have $\tilde{x} \in \mathcal{L}_\infty$, $\dot{\tilde{x}} \in \mathcal{L}_\infty$, $V_{cl} \in \mathcal{L}_\infty$, $|\bar{x}| \in \mathcal{L}_2$, and $|\mathfrak{K}^\top\mathcal{B}^\top(x_d,s)P\tilde{x}| \in \mathcal{L}_2$. This confirms that the UGES *nominal* dynamics (31) and (15) are perturbed by square integrable perturbation terms in (90) and (91). Consequently, the UGES of the closed-loop system can be concluded by Lemma 4 in [20].

To cover parametric uncertainty, an adaptive output feedback form of the control law can be obtained by replacing the unavailable states and parameters with their estimates, provided by the observer of Proposition 3.

*Corollary 5* Consider the system (1)-(2) verifying Assumptions 1-2 and 4-5, in conjunction with the observer (63)-(67). Assume that the control input is such that $\Upsilon \in \mathcal{L}_\infty$, $\Upsilon^\top\mathcal{C} \in PE$, and

$$|\bar{u}_{FI}| := \left|u_{FI}(x,\theta) - u_{FI}(\hat{z}+\Upsilon\hat{\theta},\hat{\theta})\right| \le \mathfrak{b}_0 k|\bar{z}| + \mathfrak{b}_1|\bar{\theta}|. \tag{100}$$

If the origin of the system (31) is UGES for all $K = K^\top > 0$, then there exists a sufficiently small constant $\kappa_\star$ such that the system in closed loop with $u = u_{FI}(\hat{x},\hat{\theta})$ is UGAS for all $\kappa_\star I > K > 0$.

*Proof 5* Replacing $u = u_{FI}(\hat{z}+\Upsilon\hat{\theta},\hat{\theta})$ in (23) and using (30), we get

$$\tilde{u} = -K\mathcal{B}^\top(x_d,s)P\tilde{x} + \bar{u}_{FI}, \tag{101}$$

where $\bar{u}_{FI}$ satisfies (100). Substituting the above equation in (24) yields the following tracking error dynamics

$$\dot{\tilde{x}} = [\mathcal{A}(u) - \mathcal{D} - \mathcal{B}(x_d,s)K\mathcal{B}^\top(x_d,s)P]\tilde{x} - \mathcal{B}(x_d,s)\bar{u}_{FI}, \tag{102}$$

Now, notice that (68) can be rewritten in terms of $u_d$ and $\tilde{u}$, as following

$$\dot{\bar{z}} = [\mathcal{A}_\sigma(u_d) - \mathcal{D}_\sigma]\bar{z} - \mathcal{J}_\sigma(\bar{z})\tilde{u}, \tag{103}$$

that is bounded, from (69)-(70), by

$$|\bar{z}(t)| \le \bar{z}_M := \sqrt{p_{\sigma M}/p_{\sigma m}}|\bar{z}(0)|. \tag{104}$$

Now, with similar arguments to Proof 4, it can be shown that (102)-(103) is UGES for $\bar{\theta} = 0$ and input to state stable with respect to $\bar{\theta}$. On the other hand, from Proposition 3, under Assumption 2 and $\Upsilon^\top\mathcal{C} \in PE$, we have that $\bar{\theta}$ is bounded and UGA tends to zero. As a result, UGAS of the closed-loop system can be concluded from Lemma 2.1 in [19].

## VI. EXAMPLE: SINGLE PHASE RECTIFIER

The structure of a single-phase full-bridge boost PFP is depicted in Fig. 1. From Kirchhoff's laws, the dynamic equations describing the average behavior of the converter can be obtained as follows [21]

$$L\frac{di}{dt} = -uv + v_i(t), \tag{105}$$

$$C\frac{dv}{dt} = ui - Gv, \tag{106}$$

where $i \in \mathbb{R}$ describes the current flows in the inductance $L$, and $v \in \mathbb{R}_{>0}$ is the voltage across both the capacitance $C$ and the load conductance $G$. The continuous signal $u \in [-1,1]$ operates as a control input and is fed to the PWM circuit to generate the sequence of switching positions $\delta$ its complement $\bar{\delta} := 1-\delta$. The switch position function takes values in the finite set $\{0,1\}$ and the exact model can be obtained by substituting $2\delta - 1$ as the control input. Finally,

$$v_i(t) = E\sin(\omega t), \tag{107}$$

describes the voltage of the AC input.

Assume that $i$ is an unavailable sate and $G$ is an unknown parameter in the input and output sides, respectively. Under this assumption, the control objective in the input and output sides are the unity power factor and the output voltage regulation in $V_d$, respectively. The only measurable state is $v$ and the parameters $L$, $C$, $\omega$, $E$ are known.

The system (105)-(106)can be rewritten in the form of (1)-(2) with

$$x = \begin{bmatrix} i \\ v \end{bmatrix}, \mathcal{A}_0 = 0_{2\times 2}, \mathcal{J}_1 = \begin{bmatrix} 0 & -\frac{1}{L} \\ \frac{1}{C} & 0 \end{bmatrix}, u = u, \mathcal{D} = \begin{bmatrix} 0 & 0 \\ 0 & \frac{G}{C} \end{bmatrix}, \mathcal{B}_0(s) = 0_{2\times 1}, \mathcal{E} = \begin{bmatrix} \frac{E}{L} \\ 0 \end{bmatrix}, s = \sin(\omega t),$$

$$y = v, \mathcal{C}^\top(u) = [0 \quad 1].$$

The admissible path $x_d$, $y_d$, $u_d$ can be obtained for the system by solving (3)-(4) as [21]

$$x_d = \begin{bmatrix} I_0 s \\ V_d \end{bmatrix}, y_d = V_d, u_d = (Es - LI_0 \dot{s})V_d^{-1},$$

where $I_0 = 2GV_d^2 E^{-1}$. Notice that in the precise statement of $y_d$, a sinusoidal will be added to $V_d$, which is neglected due to its zero average and small amplitude in practice. The mappings $\mathcal{J}(u)$, $\mathcal{I}(x)$, and consequently $\mathcal{A}(u)$ and $\mathcal{B}(x, s)$given by (5)-(8) are

$$\mathcal{J}(u) = \mathcal{J}_1 u = \begin{bmatrix} 0 & -\frac{1}{L} \\ \frac{1}{C} & 0 \end{bmatrix} u, \mathcal{I}(x) = \mathcal{J}_1 x = \begin{bmatrix} 0 & -\frac{1}{L} \\ \frac{1}{C} & 0 \end{bmatrix} x, \mathcal{A}(u) = \begin{bmatrix} 0 & -\frac{1}{L} \\ \frac{1}{C} & 0 \end{bmatrix} u, \mathcal{B}(x) = \begin{bmatrix} 0 & -\frac{1}{L} \\ \frac{1}{C} & 0 \end{bmatrix} x.$$

The positive definite matrix

$$P = \begin{bmatrix} L & 0 \\ 0 & C \end{bmatrix},$$

satisfies (10), (11) and (9) with

$$\mathfrak{D} = \begin{bmatrix} 0 & 0 \\ 0 & \sqrt{G} \end{bmatrix},$$

Now, the full-information controller can be formed for the system as

$$u_{FI} = u_d - K\mathcal{B}^\intercal(x_d)Px,$$

where $K$ is a constant scalar. To check the convergence condition of Proposition 1, $\mathfrak{Q}(x_d, s)$ given by (28) can be computed for the system as

$$\mathfrak{Q}(x_d)\mathfrak{Q}^\intercal(x_d) = K \begin{bmatrix} x_{1d}^2 & -x_{1d}x_{2d} \\ -x_{1d}x_{2d} & \frac{G}{K} + x_{2d}^2 \end{bmatrix},$$

which satisfies the PE condition (29) whereas it is not full rank at $s$, *i.e.*, $\omega t = n\pi$, for $n \in \mathbb{Z}$. It is straight forward to show that $\mathfrak{D}^\intercal \tilde{x} \equiv \mathcal{B}^\intercal(x_d, s)P\tilde{x} \equiv 0$ implies $\tilde{x} = 0$. Hence, $K > 0$ can be chosen arbitrarily, which is instrumental in developing the control law to the equivalent output feedback control proposed in Propositions 4 and 5.

Moving on to observer design, let

$$\Gamma(u) = \begin{bmatrix} \gamma_1 u \\ \gamma_2 \end{bmatrix},$$

which implies

$$\mathcal{A}(u) - \mathfrak{d}\mathcal{D} - \Gamma(u)\mathcal{C}^\intercal(u) = \begin{bmatrix} 0 & -\left(\frac{1}{L} + \gamma_1\right)u \\ \frac{u}{C} & -\mathfrak{d}\frac{G}{C} - \gamma_2 \end{bmatrix}.$$

By choosing $\gamma_2 > 0, \gamma_1 > -L^{-1}$, Assumption 2 holds with

$$P_o = \begin{bmatrix} \frac{L}{1 + \gamma_1 L} & 0 \\ 0 & C \end{bmatrix}, \mathfrak{D}_o = \sqrt{\mathfrak{d}G + C\gamma_2}.$$

Assumption 3 is satisfied if $u, \dot{u} \in \mathcal{L}_\infty$ and $u \in PE$ by Lemma B.2.3 in [22]. Also, Assumption 5 is established due to the specified $u_d$ for the system is being continuous and sinusoidal. This confirms that Proposition 4 is valid for the system. Under assumption $G$ is unknown, (14) of Assumption 4 holds by

$$\mathscr{b} = \frac{1}{L}\begin{bmatrix} v_i(t) \\ 0 \end{bmatrix}, \Omega(y, s) = -\frac{1}{C}\begin{bmatrix} 0 \\ y \end{bmatrix}, \theta = G.$$

### A. *Simulation results*

In this section, Simulink® of MATLAB® R2017b was utilized to conduct simulations and evaluate the control system's performance under uncertainty in $G$, $L$, $C$, and $V_d$. The simulation outcomes were obtained using system parameters and estimator-controller gains listed in Table 1 and Table 2, respectively. These system parameters were taken from the experimental setup in [23] for comparison purposes. An input source with an internal resistance of 0.02 Ω was considered, and all initial values of the system and estimator were set to zero. The PWM's switching frequency was set at 20 kHz. During these simulations, there is a 25% uncertainty observed in $G$ between the time intervals of $t = 0.05$ to 0.1 sec, in $L$ between $t = 0.15$ to 0.2 sec, and in $C$ between $t = 0.25$ to 0.3 sec. The target output throughout the test is $V_d = 200$ V, which only rises to $V_d = 210$ V between the time intervals of $t = 0.35$ to 0.4 sec.

Figures 1 to 4 depict the output voltage of the converter, the estimated input current, the estimated load resistance, and the power factor, respectively. Despite uncertainties in $G$, $L$, and $C$, the designed controller has successfully adjusted the output voltage to the desired value with good quality, as evidenced by Figures 1 and 2. Furthermore, the proposed estimator has accurately estimated the input current. The effect of changes in system parameters on the performance of the controller and estimator was insignificant.

Figure 2 shows that the estimated current is in phase with the input voltage, indicating that the control objective of achieving unity power factor has been achieved with high accuracy. This claim is supported by the time history of the power factor shown in Figure 4. Although there were small fluctuations in PF during times of parameter changes, the proper performance of the estimator and controller ensured that control and estimation objectives were met with good accuracy.

Figure 3 illustrates the estimated load conductance using the proposed estimator. The changes in $G$ at $t = 0.05\ sec$, $L$ at $t = 0.15\ sec$, and $C$ at $t = 0.25$ sec had a negligible effect on the performance of the estimator, and the estimated value quickly converged to the real value.

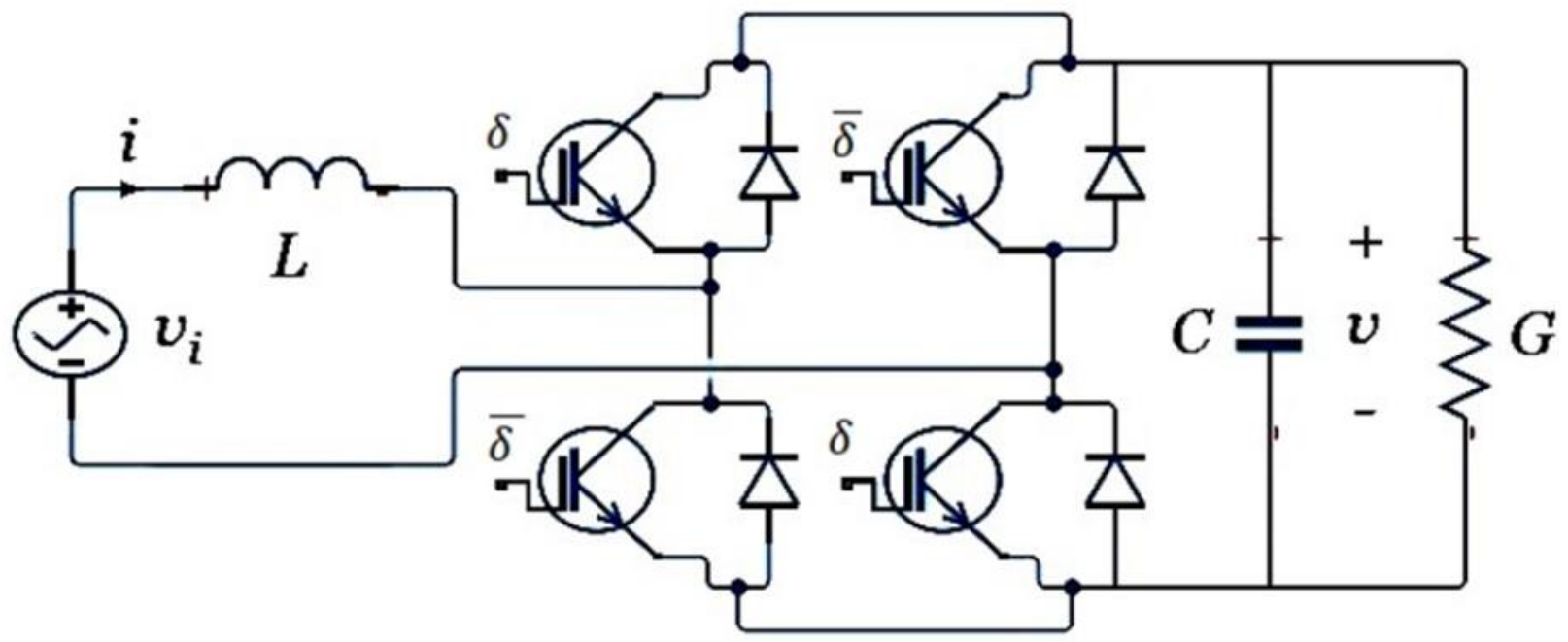

Fig. 1 A single-phase full-bridge boost PFP circuit [21].

TABLE 1
PARAMETERS OF THE SYSTEM USED IN SIMULATION

| Parameter | Value | Unit |
|---|---|---|
| $E$ | 150 | V |
| $\omega$ | $100\pi$ | rad/sec |
| $L$ | 2.13 | mH |
| $C$ | 1100 | µF |
| $G$ | 1/87 | S |

TABLE 2
PARAMETERS OF THE CONTROLLER & ESTIMATOR

| Gain | Value |
|---|---|
| $K$ | $3 \times 10^{-5}$ |
| $\gamma_1$ | $3(C^{-1} - L^{-1})$ |
| $\gamma_2$ | $2C^{-1}$ |
| $\Lambda$ | $200C$ |
| $T$ | 100 |

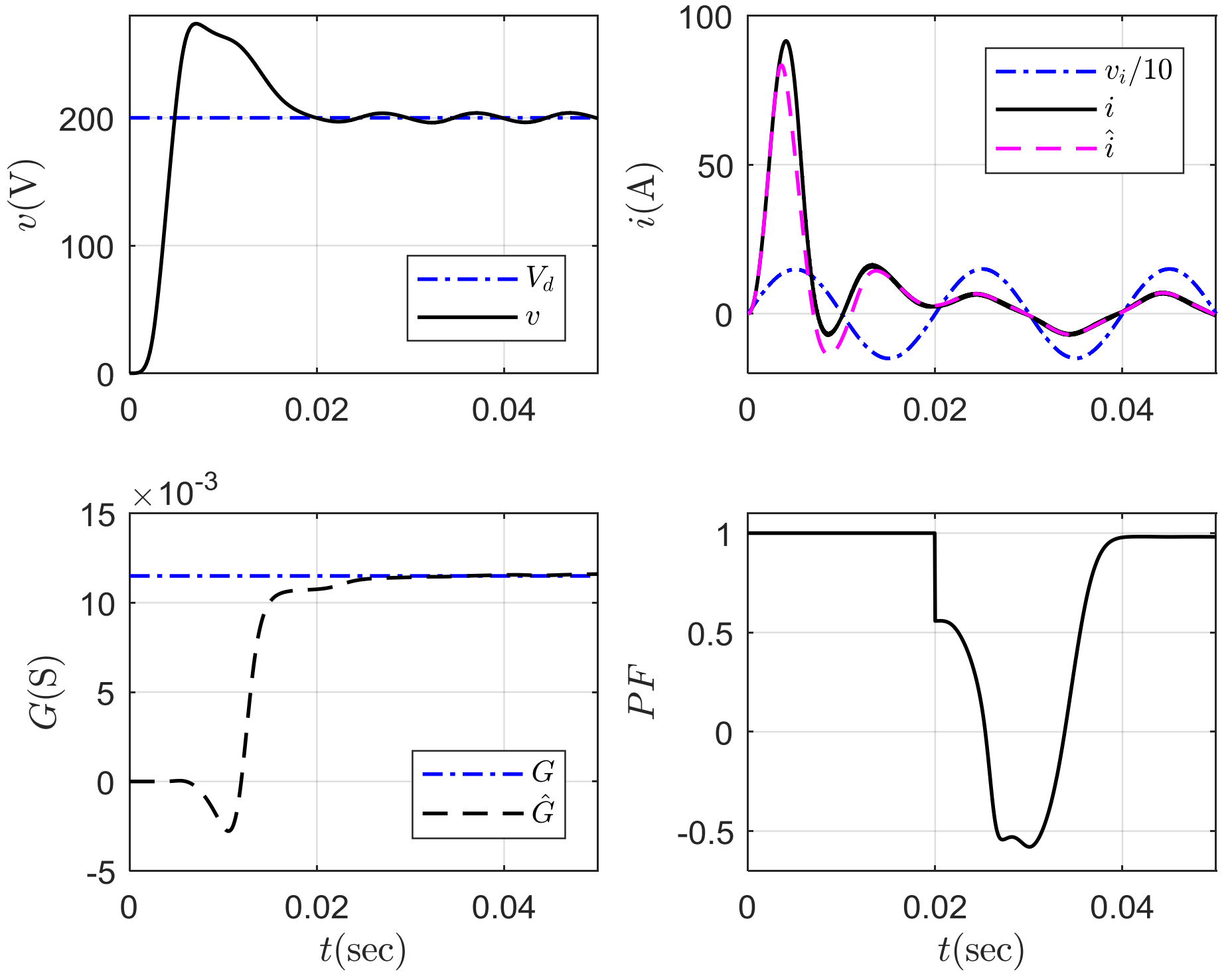

Fig. 2 Output voltage, input current and its estimate, conductance estimate, and power factor at start time.

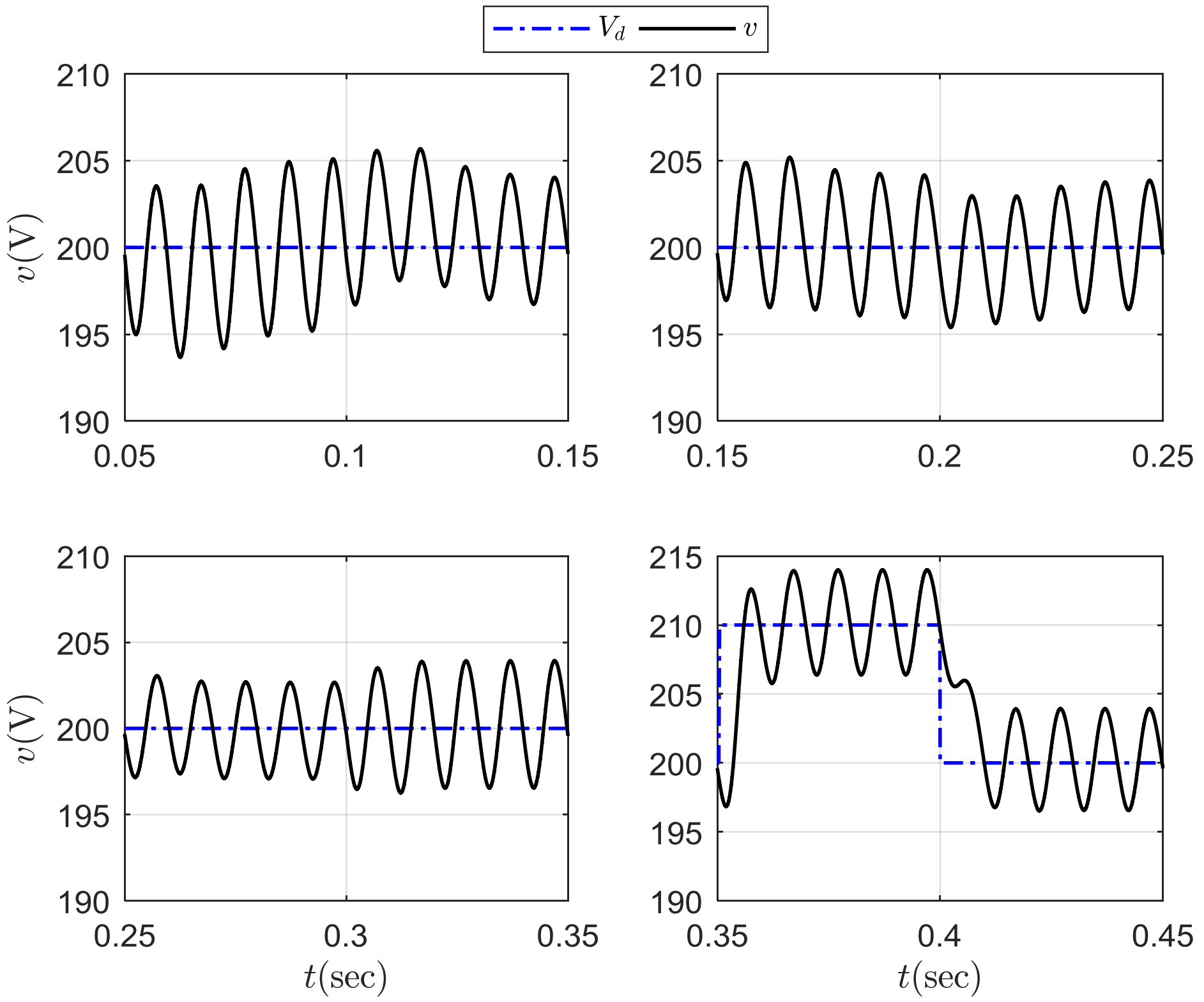

Fig. 3 Output voltage history for step changes in $G$, $L$, $C$, and $V_d$, respectively.

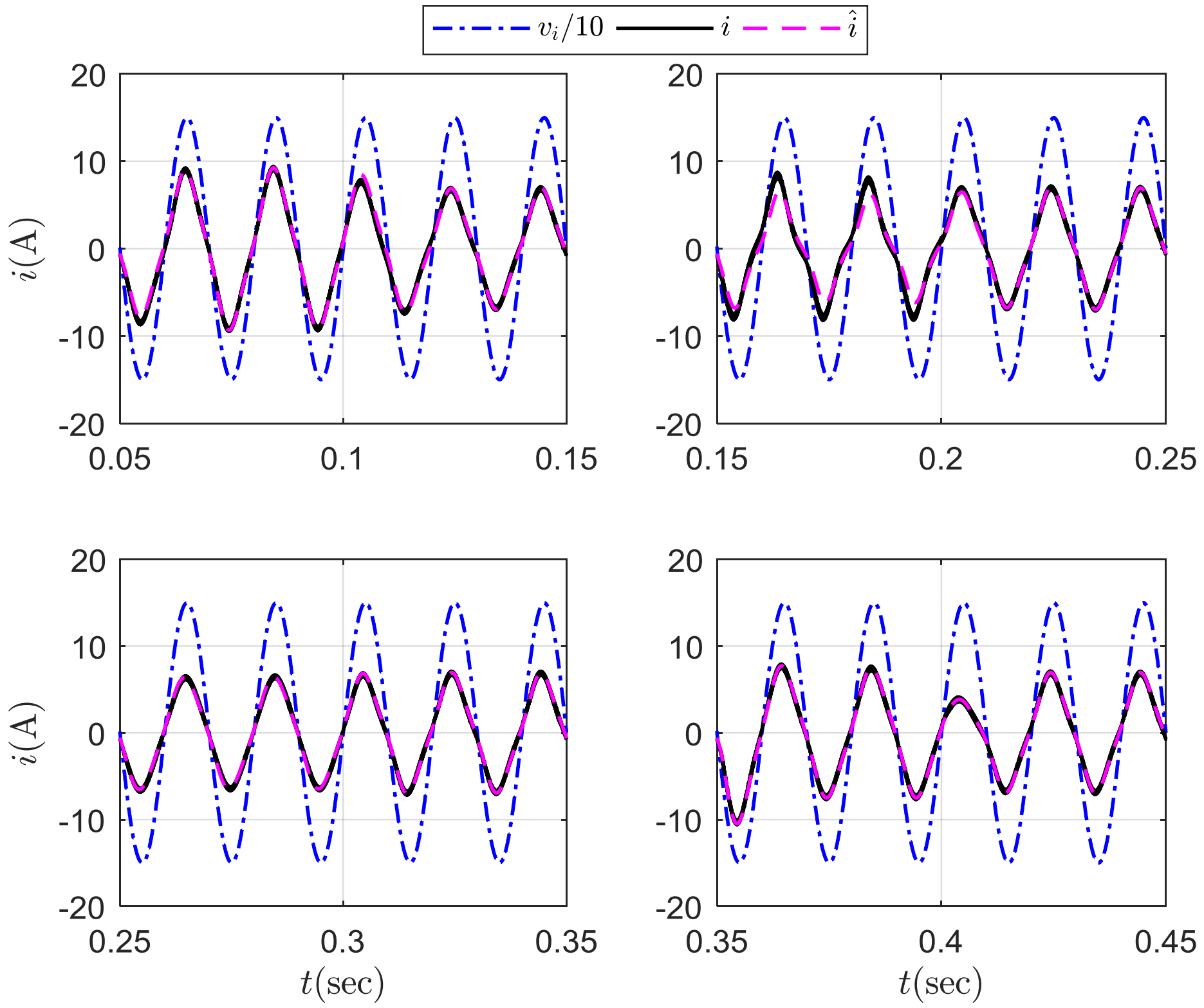

Fig. 4 Input current and its estimate histories for step changes in $G$, $L$, $C$, and $V_d$, respectively.

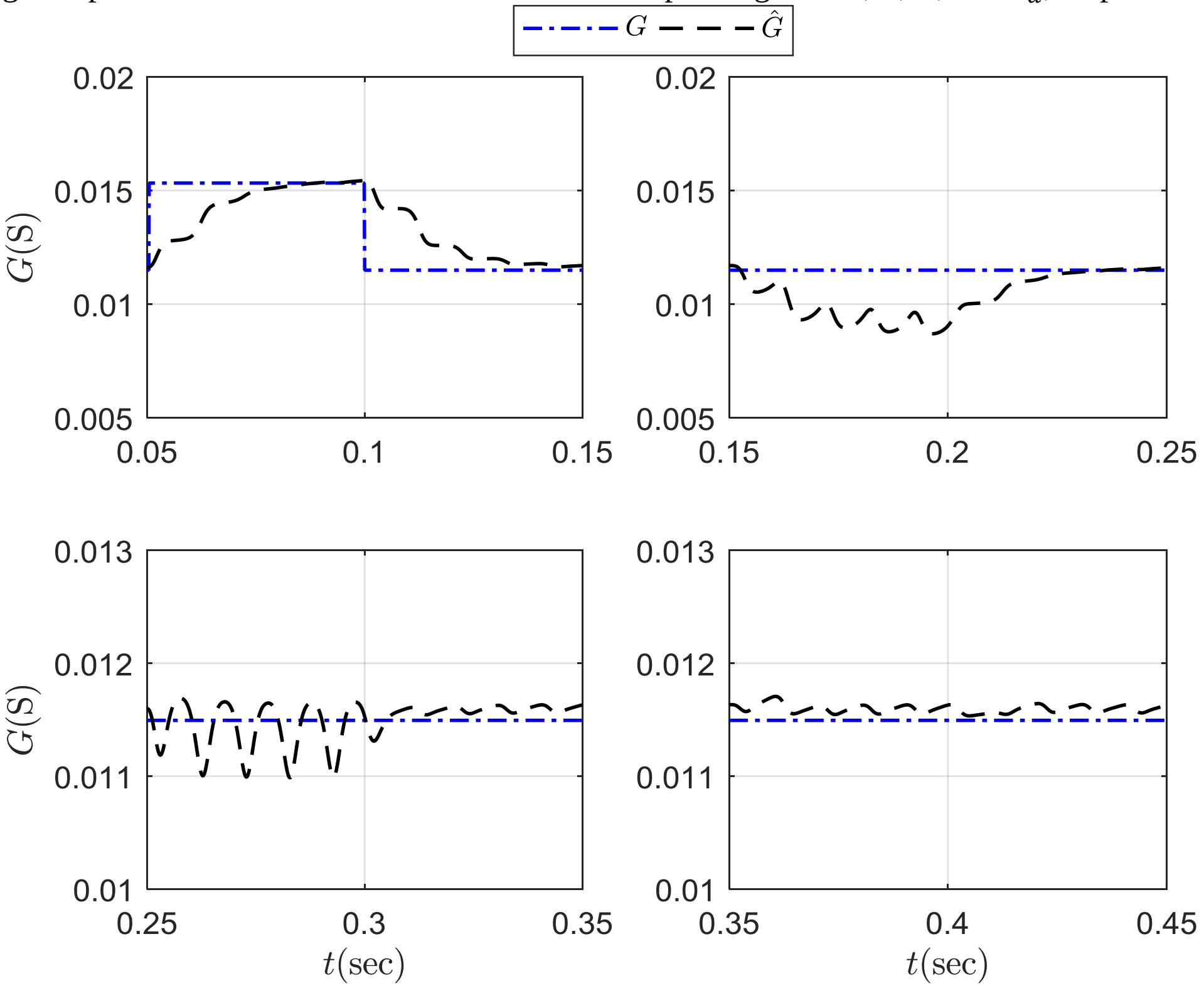

Fig. 5 Conductance estimate history for step changes in $G$, $L$, $C$, and $V_d$, respectively.

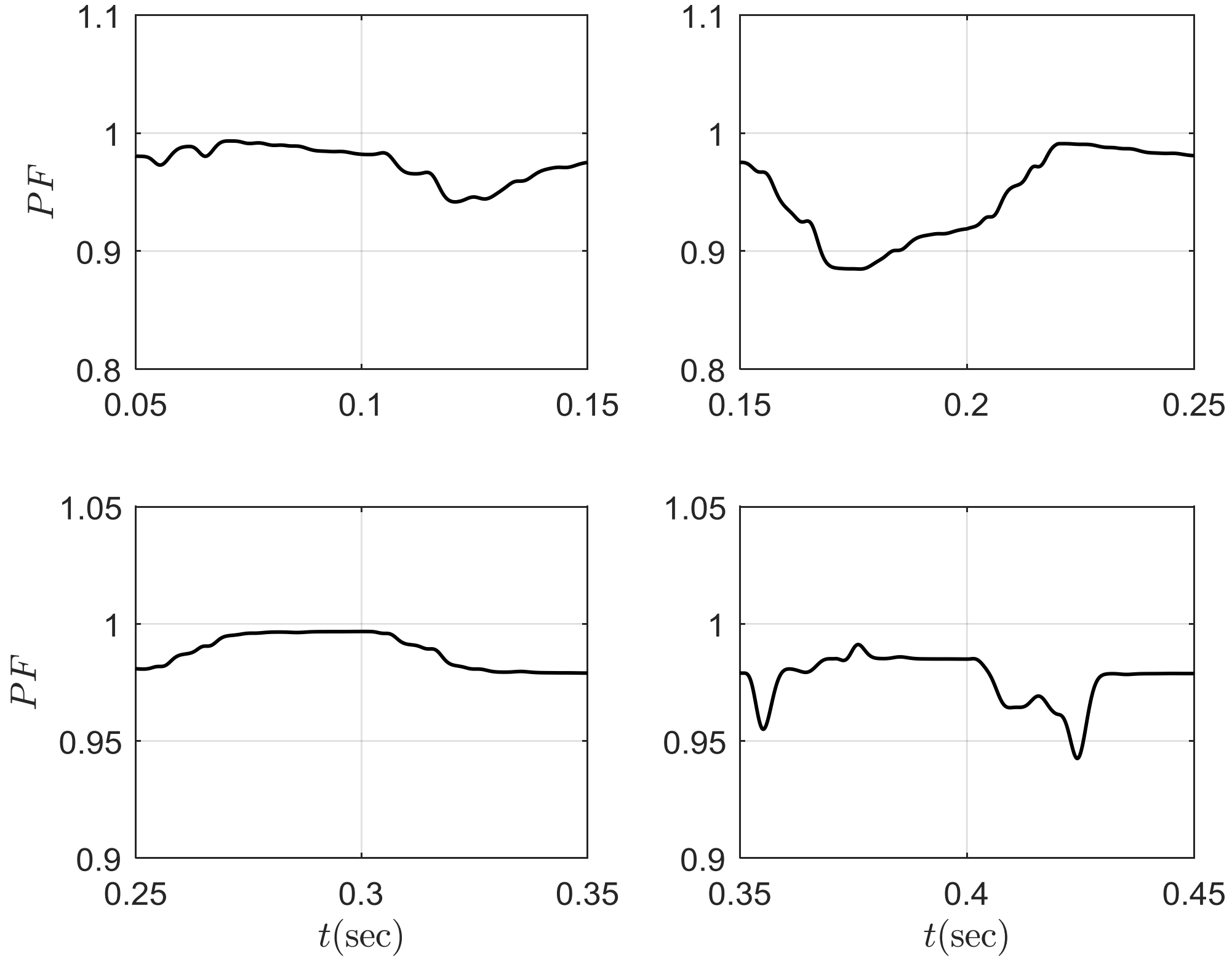

Fig. 6 Power factor history for step changes in $G$, $L$, $C$, and $V_d$, respectively.